\documentclass[12pt]{article}
\usepackage{amsmath}
\usepackage{graphicx,psfrag,epsf}
\usepackage{enumerate}
\usepackage{natbib}
\usepackage{url} 
\usepackage{amssymb}
\usepackage{booktabs,rotating,tabularx,color,multirow}
\newcommand{\blind}{1}

\begin{document}

\def\spacingset#1{\renewcommand{\baselinestretch}%
{#1}\small\normalsize} \spacingset{1}


\if1\blind
{
   \title{\bf Heterogeneous gene network estimation for single-cell transcriptomic data via a joint regularized deep neural network}
      \author{ Jingyuan Yang$^{1}$, Tao Li$^{1}$, Tianyi Wang$^{1}$, Shuangge Ma$^{2}$,
			and Mengyun Wu$^{1}$ \thanks{
  		The corresponding author, wu.mengyun@mail.shufe.edu.cn}\\
			$^{1}$  School of Statistics and Data Science, \\  Shanghai University of Finance and Economics\\
			$^{2}$  Department of Biostatistics, Yale School of Public Health}			
 \date{}
  \maketitle

} \fi

\if0\blind
{
  \bigskip
  \bigskip
  \bigskip
  \begin{center}
    {\LARGE\bf Heterogeneous gene network estimation for single-cell transcriptomic data via a joint regularized deep neural network}
\end{center}
  \medskip
} \fi

\bigskip
\begin{abstract}
Estimation of intracellular gene networks has been a critical component of single-cell transcriptomic data analysis, which can provide crucial insights into the complex interplay between genes, facilitating the discovery of the biological basis of human life at single-cell resolution. Despite notable achievements, existing methodologies often falter in their practicality, primarily due to their narrow focus on simplistic linear relationships and inadequate handling of cellular heterogeneity. To bridge these gaps, we propose a joint regularized deep neural network method incorporating Mahalanobis distance-based K-means clustering (JRDNN-KM) to estimate multiple networks for various cell subgroups simultaneously, accounting for both unknown cellular heterogeneity and zero inflation, and, more importantly, complex nonlinear relationships among genes. We introduce an innovative selection layer for network construction, along with hidden layers that include both shared and subgroup-specific neurons, to capture common patterns and subgroup-specific variations across networks. Applied to real single-cell transcriptomic data from multiple tissues and species, JRDNN-KM demonstrates higher accuracy and biological interpretability in network estimation, and more accurately identifies cell subgroups compared to current state-of-the-art methods.
Building on network construction, we further find hub genes with important biological implications and modules with statistical enrichment of biological processes.
\end{abstract}

\noindent%
\vfill

\newpage
\spacingset{1.9}  
\section{Introduction}
\label{sec:intro}
Exploring intracellular gene networks has become essential for unraveling the complex molecular systems that sustain life. These networks, involving sophisticated interactions among genes, are crucial for grasping how cells behave, regulate themselves, and how diseases arise \citep{costanzo2019global}. In recent years, the explosion of high-throughput omics data has propelled the importance of network estimation research to unprecedented heights. Among the existing techniques, the Gaussian Graphical Models (GGMs) are perhaps the most popular. In GGMs, a sparse precision matrix (the inverse of the covariance matrix) is estimated to infer the network \citep{zhang2014sparse,halama2024roadmap}, describing the conditional dependencies between genes given the rests. The estimation procedure can be simplified into a set of sparse node-wise linear regression problems with simpler optimization. Conditional dependence networks outperform correlation-based methods (where ``the other components"
are ignored) by avoiding misleading marginal dependencies and providing mechanistic insights. Crucially, they enable more accurate reconstruction of gene networks by distinguishing direct molecular interactions from indirect associations, whereas correlation-based approaches may conflate these relationships due to unaccounted system-level dependencies.

While these methods have indeed achieved notable success with bulk genomics data, they can encounter limitations due to population averaging. They also overlook the inherent diversity within cell populations, potentially leading to inaccurate network inferences. In contrast, single-cell transcriptomic data provide a unique opportunity to dissect cellular heterogeneity, allowing for more precise network reconstruction and the exploration of context-specific regulatory interactions \citep{gawad2016single}. Analyzing single-cell transcriptomic data is more challenging due to the cellular heterogeneity, dropout events, and low signal-to-noise ratio. Based on GGM, several network estimation methods have been proposed. Among them, most studies are tailored to accommodate the zero-inflated expression patterns resulting from dropout events, such as HurdleNormal \citep{mcdavid2019graphical}, PLNet \citep{xiao2022estimating}, scLGM \citep{oh2023accounting}, and PC-zinb \citep{nguyen2023structure}, and a few others take a different perspective to address the mean-correlation relationship \citep{wang2022addressing}. These methods show progress but rely on the assumption that the cells are identically and independently distributed, which is usually not true in single-cell transcriptomic data, and thus are still not practically useful.

A few other studies conduct a further step and additionally take into account the cellular heterogeneity and construct multiple networks for different cell subgroups, where the interactions, coordination, and other relationships among genes may vary. Examples include the Bayesian latent Gaussian graph mixture model (BLGGM) \citep{wu2022estimating}, which can handle the unknown cellular heterogeneity and zero inflation simultaneously. Aside from the heterogeneity of the networks, the potentially shared common structures within cell subgroups have received great attention, as cells belonging to distinct subgroups are frequently derived from the same tissue. Motivated by some studies originally developed for bulk genomics data, which are based on the fused and group Lasso penalties, a few joint network estimation methods have been proposed for effectively accommodating both the heterogeneity and homogeneity among cell subgroups, as well as the unique characteristics of single-cell transcriptomic data. Examples include the joint Gaussian copula graphical model \citep{wu2020joint}, GGM incorporating a Bayesian zero-inflated Poisson (ZIP) model \citep{dong2023joint}, and kernelized multiview signed graph learning \citep{karaaslanli2023kernelized}.

Despite considerable successes, the aforementioned GGM-based methods only focus on linear dependencies, limiting their ability to capture complex nonlinear relationships among genes that are involved in almost all biological processes. To fill this gap, a few studies take advantage of the nonparametric or distribution-free technique to detect the potential nonlinear relationships, such as the tree-based GENIE3 method \citep{huynh2010inferring}, the nonparametric test-based locCSN method \citep{wang2021constructing}, and the distribution-free-based CS-CORE method \citep{su2023cell}. However, these methods mostly adopt the unconditional construction strategy and were originally developed for single-network analysis and depend on the prior known cell subgroup information, which may still lose effectiveness under the practical scenarios with
usually complex conditional dependence and unknown cellular heterogeneity.

\begin{figure}[!ht]
	\centering
	\includegraphics[width=1\linewidth]{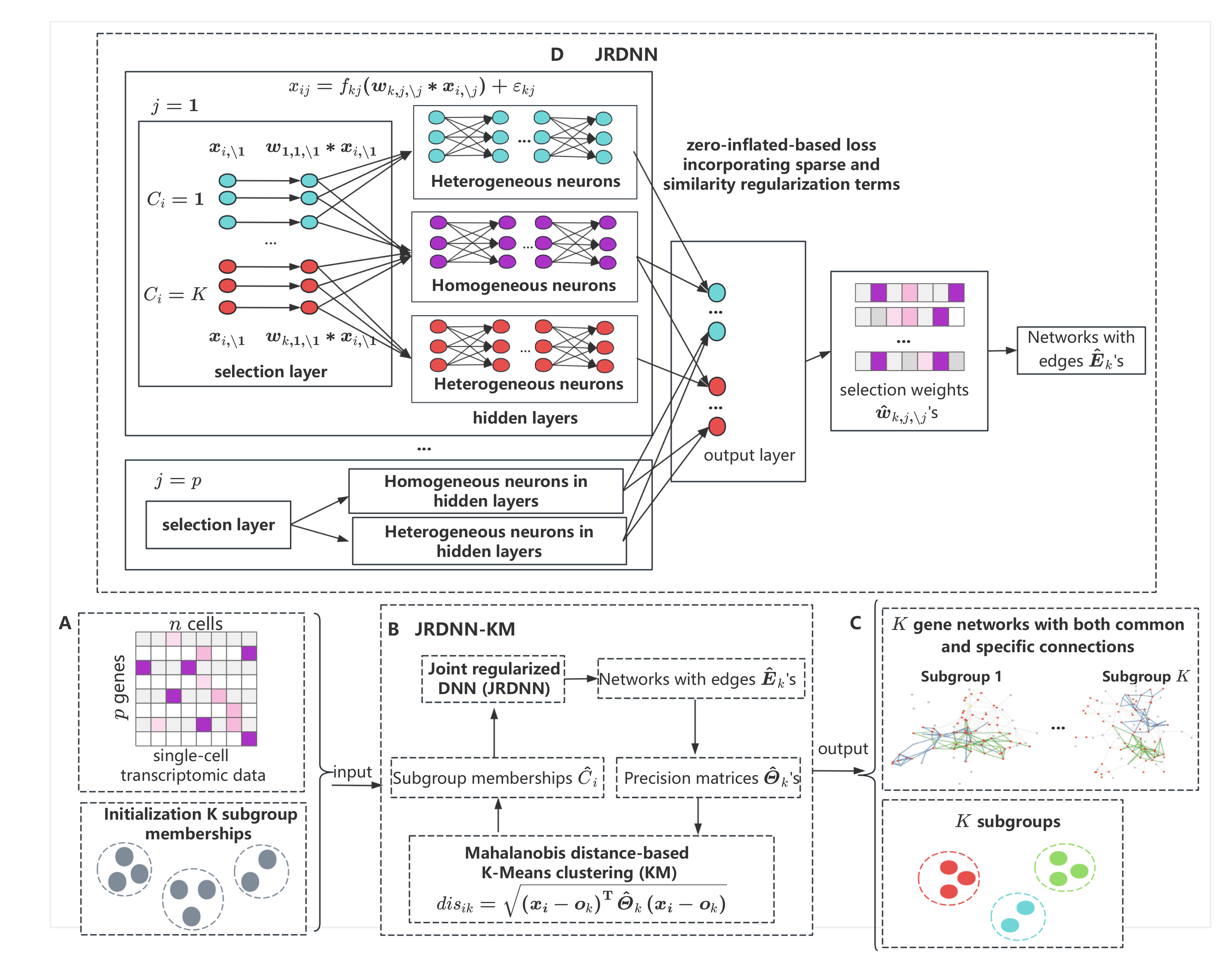}
	\caption{Workflow of JRDNN-KM. (A) Input: normalized single-cell transcriptome data $\boldsymbol{x}_i$'s and initialized cell subgroup memberships $C_i$'s. (B) JRDNN-KM: iterations between JRDNN and Mahalanobis distance-based K-means clustering. (C) Output: $K$ networks for the $K$ subgroups, with both common and specific edges, and estimated cell subgroup memberships. (D) JRDNN architecture: The architecture comprises a selection layer and hidden layers with combined homogeneous and heterogeneous neurons, and is optimized by a zero-inflated loss function that includes sparse and similarity regularization terms.}	
	\label{fig:workflow}
\end{figure}

In response to these limitations, this work develops a novel joint regularized deep neural network incorporating a Mahalanobis distance-based K-means clustering (JRDNN-KM) for single-cell transcriptomic data, achieving intracellular gene network estimation and cell subgroup identification simultaneously. A workflow of JRDNN-KM is presented in Figure \ref{fig:workflow}. Significantly advancing from the published studies, by exploiting the neural network's capacity for flexible function approximation, JRDNN-KM can capture various nonlinear dependencies and interactions among genes in different cell subgroups while concurrently handling zero-inflation. By integrating sparse and similarity regularization, and through the inclusion of both homogeneous and heterogeneous neurons in the hidden layers, we ensure that the inferred networks capture both shared and subgroup-specific structures. This enables a more comprehensive joint analysis across distinct cell subgroups. The inferred networks are further integrated into the Mahalanobis distance-based K-means clustering procedure to identify previously unknown cell subgroups. This approach offers the distinct advantage of simultaneously accounting for heterogeneity in both expression levels and network connectivity between genes. We demonstrate the performance of JRDNN-KM through extensive simulations and real single-cell transcriptomic datasets from humans and mice. JRDNN-KM detects biologically sensible cell subgroups, hub genes, and modules, significantly contributing to revealing the biological mechanisms driving cellular processes.

{
\section{Data Description}

We analyze five technically diverse single-cell transcriptomic datasets from human and mouse systems, spanning a wide spectrum of tissue types, developmental stages, and physiological contexts. Specifically, these datasets include:
\begin{itemize}
\item A human lung adenocarcinoma (LUAD) cell line \citep{tian2019benchmarking}, widely used in cancer research to investigate tumor heterogeneity;
\item Human peripheral blood mononuclear cells (PBMC) \citep{zheng2017massively}, a well-characterized immune cell reference with direct relevance to immunology and biomarker discovery;
\item Mouse embryonic stem cells (mESCs) \citep{kolodziejczyk2015single}, frequently employed in studies of early development and pluripotency;
\item Mouse liver and uterus tissues from the Mouse Cell Atlas (MCA) \citep{han2018mapping}, two distinct datasets that capture tissue-specific complexity and organ-level functions.
\end{itemize}

These five datasets exhibit substantial variation in sample size, from the smaller mESC data to the large-scale PBMC data; data quality, ranging from profiles with low zero-inflation (mESCs) to those with high dropout rates exceeding 90\% (PBMC); and subgroup distribution, including notably imbalanced cell type compositions as seen in the mouse uterus data. Each dataset is accompanied by carefully annotated cell identities based on established markers, serving as gold standards for evaluating cell subgroup detection performance. Additional details are provided in Supplementary Section S1.

Building on these rich and varied data resources, we aim to develop a novel statistical framework for heterogeneous network estimation that addresses the following two core scientific problems:
\begin{enumerate}
\item[(Q1)] How can we reliably identify cell subgroups defined by distinct patterns of gene-gene relationships, while handling the high sparsity and technical noise (e.g., dropout events) inherent in single-cell transcriptomic data?
\item[(Q2)] How can we model complex, nonlinear gene-gene relationships to reconstruct accurate and subgroup-specific gene networks and effectively explore the similarities and differences among these networks across different cell subgroups?
\end{enumerate}

The simultaneous conduct of network analysis and cell subgroup identification addresses the fundamental need to uncover previously unrecognized heterogeneity and to reveal interrelationships among molecular features across different subgroups. This practical goal directly motivates our first question (Q1), which focuses on identifying cell subgroups characterized by distinct gene-gene relationships while effectively handling technical noise. Furthermore, as articulated in the second question (Q2), moving beyond linear assumptions to model complex, nonlinear gene-gene relationships enables a more accurate representation of molecular relationships and helps to overcome the limitations of conventional Gaussian graphical models. (Q2) is further motivated by the recognition that the analysis of cellular heterogeneity should be complemented by an examination of underlying homogeneity across subgroups, particularly when cells originate from the same tissue. Such integration is essential for identifying molecular programs shared across subgroups and network alterations unique to each subgroup, thereby providing a systemic view of gene relationship modulation in different biological contexts. Together, addressing these questions establishes a powerful framework for deciphering the molecular mechanisms that underlie key processes such as development and disease pathogenesis.
}

\section{Proposed JRDNN-KM Method}
\label{sec:meth}
Suppose there are $n$ independent cells from $K$ subgroups. For the $i$th cell, denote $C_i \in \{ 1,\ldots,K\}$ as the subgroup assignment and $\boldsymbol{y}_i = \left(y_{i1},\ldots,y_{ip}\right)^{\top}$ as the $p$-dimensional vector of the observed count transcriptomic data. In practice, $\boldsymbol{y}_i$ has a high sparsity with zero-inflation due to the dropout event. Following the published studies \citep{booeshaghi2021normalization,hafemeister2019normalization}, we conduct normalization using the R package \textit{Seurat} and $\log(1+x)$ transformation on $\boldsymbol{y}_i$'s to accommodate the library size and count nature and denote the processed continuous vector as $\boldsymbol{x}_i=\left(x_{i1},\ldots,x_{ip}\right)^{\top}$.

\subsection{Network estimation based on a joint regularized deep neural network}
First, assume that the subgroup assignment is known, and for each subgroup, each gene is standardized to have a zero mean and unit variance. For notational simplicity, we preserve
$\boldsymbol{x}_i$ to represent the standardized data. {To tackle the challenges posed by the high sparsity of single-cell transcriptomic data (Q1) and the need to model nonlinear relationships in gene networks (Q2)}, when $C_i=k$, we propose a zero-inflated conditional Gaussian distribution as:
\begin{align}\label{eq:mixturedistribution}
	x_{ij}|\boldsymbol{x}_{i,\backslash j};C_i=k \sim & \mathcal{N}\left(f_{kj}(\boldsymbol{w}_{k,j,\backslash j}* \boldsymbol{x} _ { i , \backslash j }) , \sigma_{kj}^2\right)\pi_j+\delta _{0}(x_{ij})(1- \pi _{j}),
\end{align}
where $\boldsymbol{x}_{i,\backslash j}=\left(x_{i,1}, \ldots ,x_{i,j-1},x_{i,j+1}, \ldots ,x_{i,p}\right)^{\top}$, $\boldsymbol{w}_{k,j,\backslash j}=(w_{k,j1}, \ldots ,w_{k,j(j-1)},w_{k,j(j+1)}, \ldots,$ $w_{k,jp})^{\top}$ is a sparse weight vector for $\boldsymbol{x}_{i,\backslash j}$, $*$ is the element-wise product, $f_{kj}(\boldsymbol{w}_{k,j,\backslash j}*\boldsymbol{x}_{i,\backslash j })$ and $\sigma_{kj}^2$ are the mean and variance parameter of Gaussian distribution with $f_{kj}(\cdot)$ being an arbitrarily nonparametric function, $\delta _{0}(\cdot)$ is
a Dirac probability measure with a point mass at zero, and $0\leq \pi_j\leq 1$ is the probability that the $j$th gene does not express zero caused by dropouts.

Based on Model \eqref{eq:mixturedistribution}, when $C_i=k$ and $x_{ij}$ does not express zero, we have $x_{ij}=f_{kj}(\boldsymbol{w}_{k,j,\backslash j}* \boldsymbol{x} _ {i, \backslash j })+\varepsilon _ {kj}$ with $\varepsilon _ {kj}\sim \mathcal{N}\left(0,\sigma_{kj}^2\right)$, where $f_{kj}(\cdot)$ describes the nonlinear relationships between the $j$th gene and the others for the $k$th cell subgroup. Here, we introduce the sparse weight vector $\boldsymbol{w}_{k,j,\backslash j}$ to accommodate the sparse connections among genes and further construct the network. Specifically, if $w_{k,jl}\neq 0$ or $w_{k,lj}\neq 0$, the $j$th and $l$th genes have nonlinear conditional dependence and will be connected in the network. Model \eqref{eq:mixturedistribution} can be treated as a nonlinear extension of the GGM-based node-wise regression \citep{cai2016estimating, wang2019precision}.

For estimating $f_{kj}(\cdot)$ and the unknown parameters $\boldsymbol{w}_{k,j,\backslash j}$, we propose a joint regularized deep neural network (JRDNN) as shown in Figure \ref{fig:workflow}(D). Specifically, for each $j$ and $k$, the function represented by JRDNN can be written as (we omit the dependence on $j$ to simplify notation):
\begin{eqnarray}\label{DNN}
	&&g_{k}(\boldsymbol{z}_i; \boldsymbol{U}_{k} ,\boldsymbol{w}_{k})=T_{k}^{(M+1)}\left(\begin{array}{cc}
		h\circ T_{k}^{(M)}\cdots h\circ T_{k}^{(2)}\circ h \circ T_{k}^{(1)}\left(\boldsymbol{w}_{k}* \boldsymbol{z}_{i}\right)\\
		h\circ T_{0}^{(M)}\cdots h\circ T_{0}^{(2)}\circ h \circ T_{0}^{(1)}\left(\boldsymbol{w}_{k}* \boldsymbol{z}_{i}\right)
	\end{array}\right).
\end{eqnarray}
Here, $\boldsymbol{z}_i\in \mathbb{R}^{(p-1)\times 1}$ is an input vector, which is $\boldsymbol{x}_{i,\backslash j}$ in our network estimation procedure. We introduce a selection layer for $\boldsymbol{w}_k$ and consider $M$ hidden layers. Specifically, in \eqref{DNN}, we consider two sub-fully connected neural networks, consisting of heterogeneous neurons and homogeneous neurons, respectively, where heterogeneous neurons and homogeneous neurons are connected with themselves but not connected with each other. Specifically, for the $m$th hidden layer with $m=1,\cdots,M$, we consider $d_k^{(m)}$ heterogeneous neurons specific to the $k$th subgroup and $d_0^{(m)}$ homogeneous neurons shared by all $K$ subgroups, and $T_{k}^{(m)}(\boldsymbol{u})=
\boldsymbol{\Delta}^{(m)}_{k}\boldsymbol{u}+\boldsymbol{b}_k^{(m)}$ and $T_{0}^{(m)}(\boldsymbol{u})=      \boldsymbol{\Delta}^{(m)}_{0}\boldsymbol{u}+\boldsymbol{b}_0^{(m)}$ are affine transformations involving unknown parameters $\boldsymbol{\Delta}^{(m)}_{k}\in \mathbb{R}^{d_k^{(m)}\times d_k^{(m-1)}}$ and $\boldsymbol{b}_k^{(m)}\in \mathbb{R}^{d_k^{(m)}\times 1}$ and $\boldsymbol{\Delta}^{(m)}_{0}\in \mathbb{R}^{d_0^{(m)}\times d_0^{(m-1)}}$ and $\boldsymbol{b}_0^{(m)}\in \mathbb{R}^{d_0^{(m)}\times 1}$, respectively. $h(\cdot)$ is the activation function, which can be rectified linear unit (ReLU), sigmoid, tanh, and some others. In addition, an output layer with $T_{k}^{(M+1)}(\boldsymbol{u})=
\boldsymbol{\Delta}^{(M+1)}_{k}\boldsymbol{u}+\boldsymbol{b}_k^{(M+1)}$ is introduced for the output $x_{ij}$, with $\boldsymbol{\Delta}^{(M+1)}_{k}\in \mathbb{R}^{1\times \left(d_k^{(M)}+d_0^{(M)}\right)}$ and $\boldsymbol{b}_k^{(M+1)}\in \mathbb{R}^{1\times 1}$, which integrates both the heterogeneous and homogeneous neurons. Denote $\boldsymbol{U}_k$ as a vector consisting of all parameters in the $M$ hidden layers and output layer.

Based on \eqref{eq:mixturedistribution} and \eqref{DNN}, we further propose conducting joint network estimation for $K$ cell subgroups and introduce the following penalized loss function in JRDNN:
\begin{align}\label{loss}
	\nonumber l_{pn}(\boldsymbol{X};\Phi)  &= \sum_{j=1}^{p} \left\{-\sum_{i=1}^{n}\sum_{k=1}^{K}\tau_{ik}l_n(\boldsymbol{x}_{i}; \boldsymbol{U} _{kj},\boldsymbol{w}_{k,j,\backslash j}, \pi _{j})+\lambda_{1}\sum _{l \neq j}\sum _{k=1}^{K}|w_{k,jl}|\right.\\
	&\left. +  \lambda_{2}\sum _{l \neq j}\sum _{k=1}^{K}\sum_{k' \neq k} \left|1(w _{k,jl}\neq 0)-1(w _{k',jl}\neq 0)\right|+  \lambda_3 \sum_{k=1}^K|| \boldsymbol{\Delta}_{kj}^{(1)}||_F  \right\},
\end{align}
with
$ l_n(\boldsymbol{x}_{i}; \boldsymbol{U} _{kj},\boldsymbol{w}_{k,j,\backslash j}, \pi _{j}) =
\delta_{ij}\left[\log\pi_j-\frac{1}{2}\left\{\log(2\pi)+\log\left(\sigma_{kj}^2\right)+\frac{\left(
x_{ij}-g(\boldsymbol{x}_{i,\backslash j};\boldsymbol{U} _{kj},\boldsymbol{w}_{k,j,\backslash j})\right)^{2}}{\sigma_{kj}^2} \right\} \right]+(1-\delta_{ij})\log(1- \pi _{j}),
$
where $\boldsymbol{X}$ is a $n\times p$ matrix consisting of $\boldsymbol{x}_{1},\cdots, \boldsymbol{x}_{n}$, $\Phi$ denotes all unknown parameters, $1(\cdot)$ is an indicator function, $\tau_{ik}=1(C_{i}=k)$, $\delta_{ij}=1(x_{ij}\neq 0)$, and $\lambda_1$, $\lambda_2$, and $\lambda_3$ are three non-negative tuning parameters.

In \eqref{loss}, the first term is the negative log-likelihood function, which models the fit of our data. The second term, an $\ell_1$ penalty, encourages sparsity in the selection layer parameters $w_{k,jl}$'s to facilitate sparse network estimation. {The third and fourth terms model the commonality among cell subgroups, thereby addressing the need to incorporate inter-subgroup network similarities and differences in (Q2).} Specifically, for any pairs of $j$ and $l$, the third term promotes $1(w _{k,jl}\neq 0)=1(w _{k',jl}\neq 0)$, resulting in that $w _{k,jl}$ and $w _{k',jl}$ tend to be zero or nonzero simultaneously. Thus, the $K$ subgroups potentially share some common edges. We further introduce the fourth term to accommodate the similarity in connection strength of the edges. Specifically, the fourth term imposes the F-norm on the weights $\boldsymbol{\Delta}^{(1)}_{k}$ of heterogeneous neurons in the first layer, promoting the elements of $\boldsymbol{\Delta}^{(1)}_{k}$ shrunk towards zero simultaneously and thus pruning all heterogeneous neurons in the $M$ hidden layers. With \eqref{loss}, the constructed networks of different cell subgroups not only reflect heterogeneity but also exhibit common connections with similar strength.

\subsection{Heterogeneous Analysis}
In practice, the subgroup assignments $C_i$'s are not always observed. {
To address the unknown cell heterogeneity in (Q1), we propose using K-means clustering with Mahalanobis distance,} which is defined as $ dis_{ik}= \sqrt{\left(\boldsymbol{x_i} - \boldsymbol{o}_k\right)^{\top}\boldsymbol{\Sigma}_k^{-1}\left(\boldsymbol{x_i} - \boldsymbol{o}_k\right)}$, where $\boldsymbol{o}_k$ is the centroid of the $k$th subgroup and $\boldsymbol{\Sigma}_k$ is the covariance matrix for the $p$ genes. Instead of estimating $\boldsymbol{\Sigma}_k$, we consider estimating the precision matrix $\boldsymbol{\Theta}_k=\boldsymbol{\Sigma}_k^{-1}$ (which describes the conditional dependence between any two genes given the rest) directly with the network constructed using JRDNN. Specifically, $\hat{\boldsymbol{\Theta}}_{k}=(\hat{\Theta}_{k,jl})_{p \times p}$ is estimated with $\hat{\Theta}_{k,jl}=\hat{\Theta}_{k,lj}=1$ if $(j, l) \in \hat{E}_k$ and $\hat{\Theta}_{k,jl}=\hat{\Theta}_{k,lj}=0$ otherwise, for $j\neq l$, and $\hat{\Theta}_{k,ll}=1$, where $\hat{E}_k=\left\{(j,l): \hat{w}_{k, jl}\neq 0\textrm{ or }\hat{w}_{k, lj}\neq 0\right\}$ with $\hat{w}_{k, jl}$ being estimated using JRDNN.

Here, binary $\Theta_{k,jl}$ acts solely as a dependency indicator, not a magnitude measure. This design aligns with Mahalanobis distance-based K-means clustering: K-means partitions data based on relative dissimilarity, while the Mahalanobis distance incorporating the precision matrix's sparsity pattern discriminates local dependence from independence. This adaptation captures direct conditional relationships, ensuring clusters reflect similarity in the underlying graph-defined dependency network. The strategy accounts for heterogeneity in both gene expression levels and gene-gene relationships, thereby enabling more accurate cell subgroup identification.

\subsection{Computation}
To obtain the final network estimation and cell subgroup identification results, we implement an iterative strategy (Supplementary Algorithm S1) that alternates between JRDNN-based joint network estimation and Mahalanobis distance-based K-means clustering until convergence. Each iteration proceeds as follows: First, using current subgroup memberships
$C_i$ derived from K-means clustering, we perform JRDNN estimation. Consistent with established practice for focusing on variable relationships rather than mean differences \citep{ma2016joint}, we standardize the data within each subgroup during this step. The updated network estimate $\hat{\Theta}_k$ is then fed back into the K-means algorithm, which clusters the unstandardized data using this current estimate to compute Mahalanobis distances. For the JRDNN optimization, we minimize the penalized loss function (\ref{loss}) using stochastic gradient descent (SGD; Supplementary Algorithm S2) implemented via Python's PyTorch framework. To address the discontinuity of the indicator function in (\ref{loss}), we approximate
$1(w _{k,jl}\neq 0)$ with the smooth function $1-e^{-\frac{w _{k,jl}^2}{\xi}}$, where
$\xi$ is a small positive constant and controls the approximation fidelity. The K-means clustering step employs our specialized algorithm (Supplementary Algorithm S3) that iteratively updates cluster centroids and reassigns memberships based on Mahalanobis distance. The complete parameter settings for all steps, examination of the memory and time usage of the proposed algorithm, and discussions on the convergence of the two-step iterative optimization are provided in Supplementary Section S2.

\section{Application}

{We apply the proposed JRDNN-KM method to the five single-cell transcriptomic datasets in Section 2 and address questions (Q1)-(Q2).} First, to control the quality of datasets, we follow the published studies and remove cells expressing less than 500 genes and genes expressed in less than 100 cells. All data are then subjected to batch effect correction using MNN \citep{haghverdi2018batch}, normalized for library size using the total count normalization (TC) method \citep{cole2019performance}, and filtered for 100 highly variable (HV) genes using the variance stabilizing transformation (vst) method in R package ``Seruat'' \citep{hafemeister2019normalization}.

{ Before further analysis, we perform a series of model checks for each cell type based on the gold-standard labels. This includes validating the JRDNN model's capture of non-linear gene relationships through residual correlation analysis, and verifying the normality of residuals and the absence of outliers. As detailed in Section S3 of the Supplementary Materials, these checks confirm that the JRDNN method is appropriate for all five datasets.}

In addition to the proposed JRDNN-KM, we consider ten competing methods, including BLGGM \citep{wu2022estimating}, JGNsc \citep{dong2023joint}, JSEM \citep{ma2016joint}, SpQN \citep{wang2022addressing}, Normalisr \citep{wang2021single}, GENIE3 \citep{huynh2010inferring}, locCSN \citep{wang2021constructing}, CS-CORE \citep{su2023cell}, SC3 \citep{kiselev2017sc3}, and Seurat \citep{butler2018integrating}. Details for these competing methods and their implementation are provided in Section S4 of the Supplementary Materials.

\subsection{JRDNN-KM leads to biologically sensible cell subgroups}

We first consider the candidate sequence $\{2,3,\cdots,8,9,10\}$ for $K$. For JRDNN-KM, the silhouette coefficient identifies three subgroups for all the LUAD, PBMC, and mESCs datasets, and five and six subgroups for the mouse liver and uterus datasets, respectively. The true numbers of cell subgroups are all identified correctly. For making a fair comparison, in the following analysis, $K$ is set as the true number of cell subgroups for all methods, which has been usually considered in recent single-cell transcriptomic data analysis \citep{li2023zinbmm}.

\begin{figure}[!ht]
	\centering
	\includegraphics[scale=1]{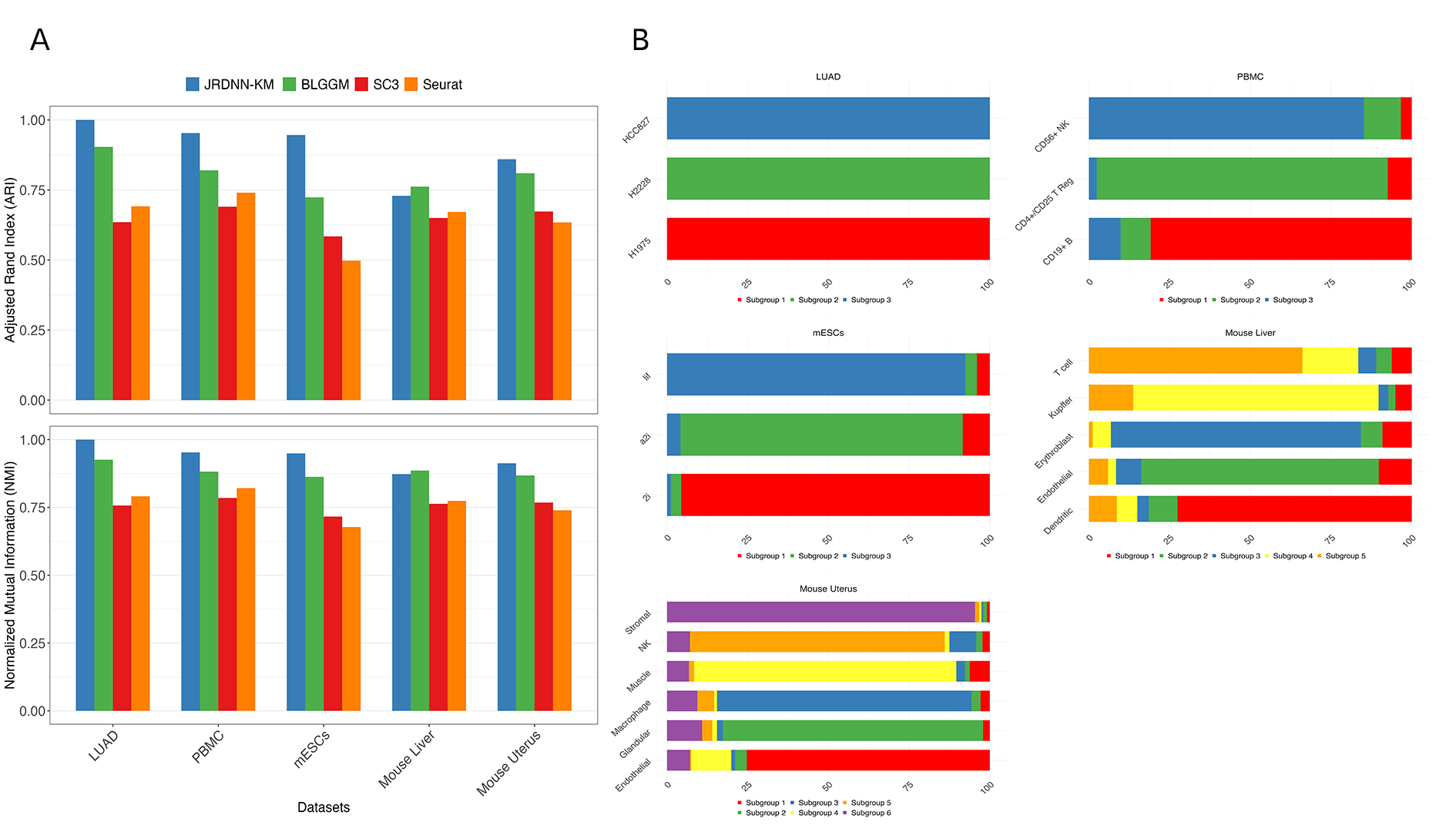}
	\caption{Heterogeneity analysis results on five real single-cell transcriptomic datasets. (A) ARI and NMI values with different methods. (B) Proportions of identified cell subgroups with JRDNN-KM in the true cell subgroups.}
	\label{fig:ARI}
\end{figure}

In Figure \ref{fig:ARI}(A), the Adjusted Rand Index (ARI) and Normalized Mutual Information (NMI) are used to evaluate subgroup identification performance. Both metrics range from 0 to 1, with higher values indicating greater accuracy. Note that ARI and NMI values are not available for JGNsc, JSEM, SpQN, Normalisr, GENIE3, locCSN, and CS-CORE, as these methods are designed for network inference and do not identify unknown cell subgroups. The proportions of cell subgroups identified by JRDNN-KM that correspond to the true subgroups are shown in Figure \ref{fig:ARI}(B), with results for alternative methods provided in Supplementary Figure S4.

Overall, JRDNN-KM demonstrates superior or competitive performance across multiple datasets, achieving satisfactory ARI and NMI values. { This indicates that the proposed strategy effectively addresses (Q1), yielding biologically meaningful cell subgroups characterized by heterogeneous gene networks.} Specifically, in the LUAD dataset, JRDNN-KM achieves both ARI and NMI values of 1.00. This performance markedly outstrips other alternative methods. Additionally, JRDNN-KM maintains higher performance metrics across diverse cellular contexts, including PBMC, mESCs, and mouse uterus cells, where the mouse uterus dataset has a highly imbalanced sample distribution with some subgroups having a small number of cells.  For the mouse liver cell data, which has a high dropout rate (90.13\%) and small sample size, JRDNN-KM performs slightly worse than BLGMM, but still behaves much better than SC3 and Seurat. Although there is a certain degree of mis-identification for some datasets, Figure \ref{fig:ARI}(B) shows that each true subgroup is dominated by one of the subgroups identified with JRDNN-KM, further validating its utility in precision oncology and regenerative medicine. The two methods, SC3 and Seurat, that do not accommodate the network structure among genes, always have inferior cell subgroup identification accuracy.

\subsection{JRDNN-KM demonstrates high-accuracy alignment with well-established biological networks}

To validate the accuracy of network reconstruction, we employ established biological network databases following the published studies \citep{chevalley2025large,pratapa2020benchmarking}. Specifically, we utilize three major databases: CORUM for protein complexes, STRING for protein-protein interactions (PPIs), and PerturbAtlas for genetic and chemical perturbations. Within STRING, we distinguish between two interaction types: PPI(F), which encompasses functional interactions, and PPI(P), which contains experimentally verified physical interactions. Given that the cell-type-specific information in these databases remains relatively limited, similar to many existing studies, we construct non-cell-type-specific networks here, which remain consistent across different cell types. For each of these four reference databases, we extract subnetworks by matching our 100 HV genes, with the resulting edge counts documented in Supplementary Table S3. Details regarding network construction and database specifications are available in Supplementary Section S5.

\begin{figure}[!ht]
\centering
\includegraphics[width=\textwidth, keepaspectratio]{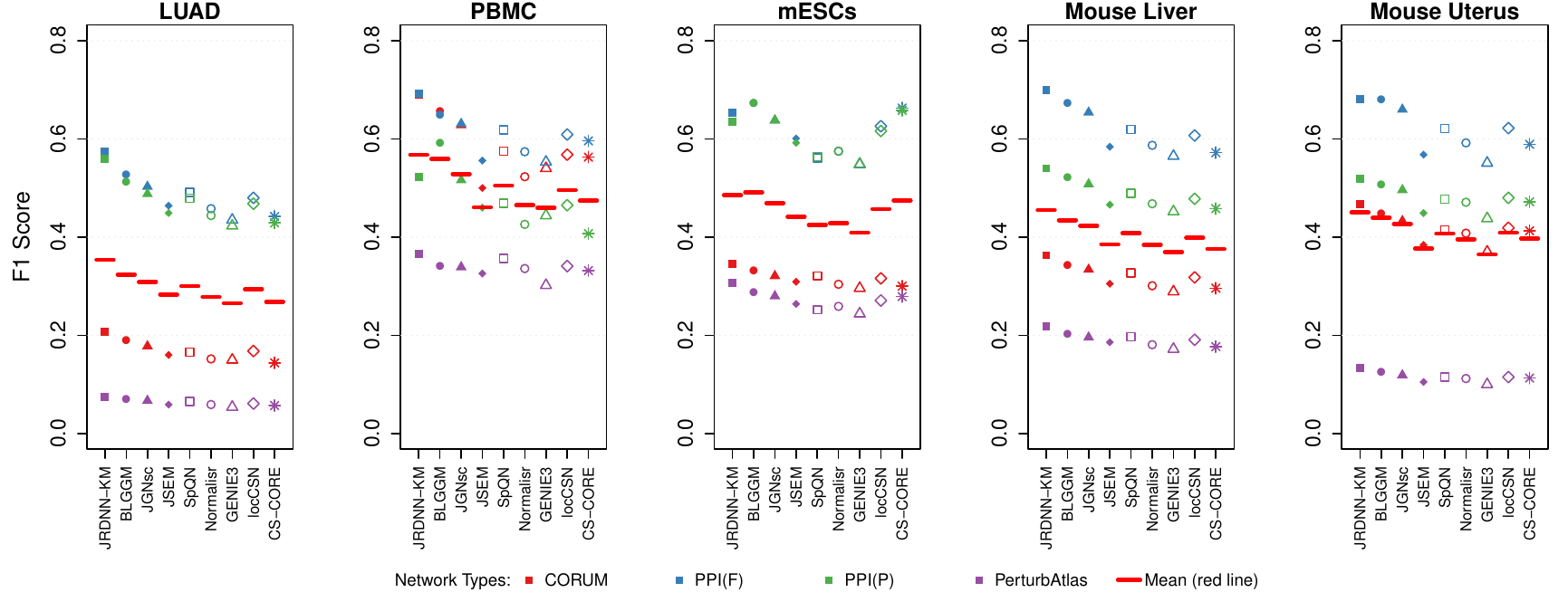}
\caption{
F1 scores of different methods evaluated against four reference networks across five datasets. Shapes represent methods, colors indicate reference networks, and horizontal lines show mean values across the four networks.
}
\label{fig:network_comparison}
\end{figure}

We evaluate network estimation performance using Recall, Precision, and F1 score by comparing the estimated edges against those from four established reference networks. F1 scores are summarized in Figure \ref{fig:network_comparison}, while detailed Recall and Precision values are provided in Supplementary Figures S5 and S6. For the seven competing methods that require prior subgroup information, we supply the true subgroup memberships. Note that SC3 and Seurat are excluded from this comparison as they do not perform network estimation. The results show that our method achieves the highest F1 scores across all five datasets when compared to the CORUM-based and PerturbAtlas-based networks. Owing to the larger number of edges included in the PerturbAtlas-based network, which can lead to reduced recall, all methods exhibit relatively lower F1 scores in this context. For the two PPI networks, our method also delivers satisfactory performance. It attains the highest F1 scores in both PPI networks across the LUAD, mouse liver, and mouse uterus datasets. In the PBMC dataset with a high zero-inflation rate and the mESCs dataset with a limited sample size, our results on either PPI (F) or PPI (P) are slightly lower than those of BLGGM but still exceed the performance of other methods. In terms of average performance across all four networks, our method ranks first in four out of the five datasets (LUAD, PBMC, mouse liver, and mouse uterus) and second in the mESCs dataset, with performance nearly matching that of BLGGM. {These findings validate the capability of the proposed network reconstruction framework in effectively addressing (Q2).}

\subsection{JRDNN-KM constructs heterogeneous networks with common edges}
\begin{figure*}[!ht]
	\centering
	\includegraphics[scale=0.6]{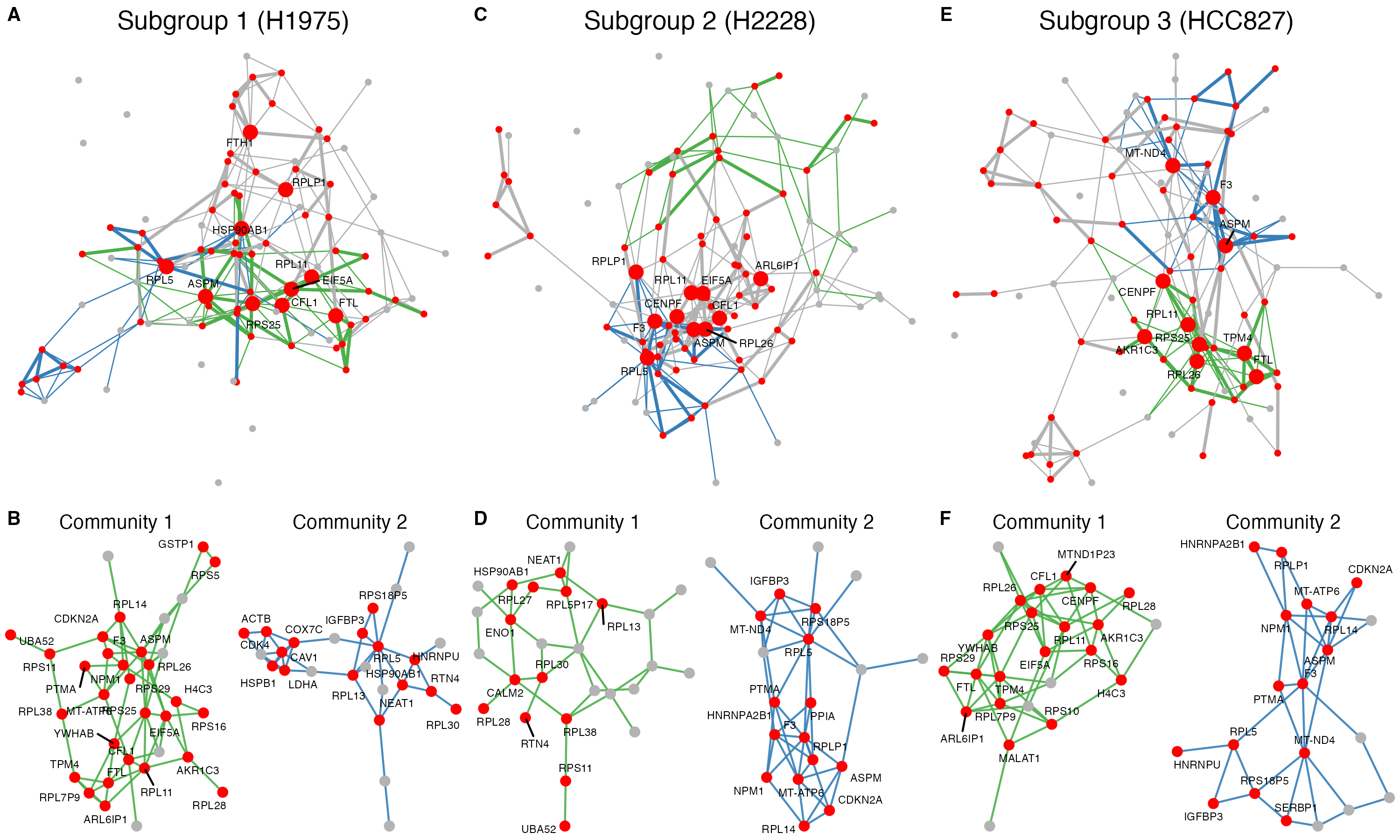}
	\caption{Networks constructed with JRDNN-KM for subgroup 1 (A), subgroup 2 (C), and subgroup 3 (E) of the LUAD dataset, where the common connections across all subgroups are highlighted by thick edges with involved genes highlighted with red color, and the hub genes with the top ten largest degrees are highlighted with a bigger point. (B), (D), and (F): Two representative communities detected using the Louvain algorithm for subgroup 1, subgroup 2, and subgroup 3, respectively.}
	\label{fig:LUAD-network}
\end{figure*}

The inferred networks with JRDNN-KM are shown in Figure \ref{fig:LUAD-network} for the LUAD dataset and Supplementary Figures S7-S10 for the rest four datasets. The identified numbers of edges for different subgroups are reported in Supplementary Figures S11-S15, where 70, 88, 76, 2, and 0 common edges are shared by all subgroups for the LUAD, PBMC, mESCs, and mouse liver and uterus datasets, respectively. The numbers of common edges for the mouse liver and uterus datasets are smaller compared to the other three datasets. This is reasonable, as the corresponding cell subgroups have relatively larger diversities. We take a closer look at the Kupffer  and T cell subgroups for the mouse liver dataset, which both participate in the regulation process of the immune system, and identify 71 common edges. For the mouse uterus dataset, we also examine the two immune regulation-related cell subgroups, Macrophage and NK cells, and identifies 26 common edges. In addition, 28 common edges are shared by the Endothelial and Muscle cells, which both participate in maintaining the normal physiological functions of the body. Comparative results across methods are provided in Supplementary Figures S11-S15. It can be seen that the nine methods identify a moderate number of overlapping edges, where the overlapping ratio of the identified edges between the proposed method and others lies in [52.7\%,82.8\%], and JRDNN-KM can effectively exploit the common information across different subgroups.

\subsection{Investigation of hub genes for the LUAD  data}
We further take the LUAD data from humans as an example and investigate the hub genes in the inferred networks with JRDNN-KM. The hub genes are important topological features of networks that usually have functional relevance. In this study, we consider the top ten genes with the largest degrees in the networks (shown in Supplementary Table S4 and highlighted with a bigger point in Figure \ref{fig:LUAD-network} and Supplementary Figures S7-S10). It can be seen that JRDNN-KM identifies two common hub genes (RPR11 and ASPM) shared by all three cell lines (H1975, H2228, and HCC827) and nine common hub genes shared by two of the three cell lines. The similarity across these three cell lines has been universally recognized in the literature. In addition, two specific hub genes (FTH1 and HSP90AB1), one specific gene (ARL6IP1), and three specific genes (MT-ND4, AKR1C3, and TPM4) are identified for subgroup 1 (H1975), subgroup 2 (H2228), and subgroup 3 (HCC827), respectively.

The top ten hub genes identified with the alternatives and the comparison results among different methods are also reported in Supplementary Table S4. It is observed that JRDNN-KM detects a few well-known marker genes that are found by the majority of approaches and also reveals some promising novel findings. Specifically, the two common hub genes, RPL11 and ASPM, shared by all three cell lines, are also recognized by most of the competing methods. Here, ASPM has emerged as a critical player in cancer biology, specifically influencing cancer aggressiveness and stemness. ASPM's overexpression in various malignancies has been demonstrated to be strongly associated with poor patient outcomes, underscoring its role in oncogenic processes and its impact on cancer cell behavior and tumor evolution. In addition, RPL11 is crucial in cancer progression through its interaction with the MDM2-P53 pathway. In particular, RPL11 acts as a tumor suppressor by stabilizing P53, particularly when PICT1 is deficient with the inhibition of MDM2's activity, which correlates with slower tumor growth and potentially improved patient outcomes.

We further examine the subgroup-specific hub genes, where JRDNN-KM demonstrates its unique capability by recognizing hub genes that most competing methods missed. For instance, in the H1975 cell line, the FTL gene and in the HCC827 cell line, the AKR1C3 gene are only identified by JRDNN-KM and locCSN. These methods, both employing non-linear approaches, effectively highlight the non-linear regulatory relationships among certain genes in LUAD, which might be overlooked before. Additionally, the CFL1 gene in the H2228 cell line is recognized only by JRDNN-KM. Biological implications of these genes have been well recognized in the literature. Specifically, the FTL gene plays a pivotal role in regulating ferroptosis by controlling iron storage and reducing oxidative stress under the governance of the NRF2 pathway. This regulation helps LUAD cells evade ferroptosis, thereby contributing to potential treatment resistance. AKR1C3 is consistently overexpressed in tumor tissues, suggesting its significant association with the disease and indicating its involvement in tumor progression and resistance to erlotinib therapy in LUAD. In addition, the CFL1 gene and its functional gene network have been suggested as prognostic biomarkers for lung adenocarcinoma, which can also guide chemotherapeutic interventions. References supporting the discussions on biological functionalities are provided in Supplementary Section S5.

\subsection{Community detection and GO enrichment analysis for the LUAD data}

We continue to conduct community (module) detection on the referred networks using the Louvain algorithm and show two communities with the largest sizes in Figure \ref{fig:LUAD-network} and Supplementary Figures S7-S10. Both homogeneous and heterogeneous edges are observed in these communities. Similar to the analysis on the hub genes, a deeper examination based on the gene ontology (GO) enrichment analysis is conducted for the LUAD data, where the top five GO terms with the smallest P-values for each subgroup are presented in Supplementary Figure S16. The results suggest that the communities share some significantly enriched GO terms, indicating functional and biological connections among the related genes. The community detection analysis is also conducted for the competing methods, and the top ten significant GO terms involved in the two largest communities are reported in Supplementary Tables S5-S7. It can be seen that some GO terms are shared by all the nine methods. Compared to the H1975 and H2228 cell lines, JRDNN-KM has more unique findings in the HCC827 cell line.

Specifically, as shown in Supplementary Figure S16, ``cytoplasmic translation'' is among the top five significant GO terms in all three subgroups. This biological process has been found to be crucial as it facilitates the synthesis of proteins necessary for cancer cell growth and proliferation and is notably involved in the progression of LUAD. In addition, the p53 class mediator-related GO terms are significantly enriched in both H2228 and HCC827 cell lines. Although not featured among the top five significant terms, these terms are also significant in the H1975 cell line, specifically the ``regulation of signal transduction by p53 class mediator'' with a p-value of $4.736 \times10^{-4}$, and ``positive regulation of signal transduction by p53 class mediator'' with a p-value of $ 7.894\times10^{-4}$. In LUAD, p53 has been shown to suppress tumor development by promoting the differentiation of alveolar type 1 (AT1) cells from transitional states during alveolar repair, effectively governing cell state to prevent the persistence of cancerous cell types and maintain lung tissue integrity. These two terms are also detected by all competing methods with significant p-values.

Similar to the analysis of hub genes, JRDNN-KM identifies some unique GO terms, such as ``regulation of translation'' found only by JRDNN-KM and locCSN in the H1975 cell line and ``regulation of protein modification'' discovered only by JRDNN-KM in the HCC827 cell line, suggesting that JRDNN-KM reveals relevant novel biological processes. Recent research shows that in the H1975 cell line, significant alterations in translation regulators were observed, particularly in rociletinib-resistant cells. These changes may affect the synthesis of key proteins, potentially influencing tumor progression and response to therapy. And in HCC827 cells, inhibiting N-linked glycosylation, a key protein modification, disrupts EGFR function, reducing proliferation and inducing senescence. These unique findings of JRDNN-KM correlate well with practical biological explanations, providing valuable insights for research and clinical applications (References supporting these discussions are provided in Supplementary Section S5).

{
\subsection{Sensitivity and Subsampling Analysis}

For the five datasets, we further investigate the influence of DNN hyperparameters on the analytical outcomes of the JRDNN-KM method. Specifically, in addition to the two hidden layers and the corresponding network width (baseline width) used in the prior analysis, we also consider alternative configurations of depth (4 and 7 layers) and width (0.5$\times$ and 2$\times$ the baseline width). The results for cell subgroup identification and network estimation under these configurations are presented in Supplementary Figure S17. We find that the baseline width yields optimal performance compared to other width settings. Regarding network depth, while accuracy improves modestly with more layers, the marginal gains are limited. Therefore, to balance performance improvement with computational cost, the two-layer architecture is adopted as the rational choice in our method.

Moreover, to evaluate the robustness of our model, we perform a subsampling analysis across the five real datasets. Specifically, we randomly subsample 80\% of the data from each dataset and repeat this process 50 times to create independent replicates. As shown in Supplementary Figures S18-S21, our proposed method consistently maintains its performance advantage and exhibits high stability across the subsampled replicates.

}

{
\section{Application-Based Simulation}
\label{sec:simulation}

In this section, we provide application-based simulations to evaluate our method using three benchmark datasets: the Gonadal Sex Determination (GSD) dataset \citep{pratapa2020benchmarking}, scMultiSim-T3 dataset \citep{li2025scmultisim}, and SERGIO-DS1 dataset \citep{dibaeinia2020sergio}. These datasets were derived from real biological networks and incorporate practical biological processes, offering a challenging and biologically realistic benchmark for testing network inference algorithms. Specifically, GSD dataset is based on a Boolean network model of gonadal sex determination, which captures the bipotential differentiation of gonads into either male (Sertoli cells) or female (Granulosa cells) fates. scMultiSim-T3 dataset is generated using scMultiSim, a multi-modality single-cell data simulator that integrates biological factors including gene regulatory networks (GRNs) and cell-cell interactions (CCIs). SERGIO-DS1 dataset is generated by the single-cell gene expression simulator SERGIO, which is constructed based on a GRN derived from Ecoli. More detailed descriptions of these datasets are provided in Supplementary Section S6.

Similar to real data analysis, we report the summarized ARI, F1, Recall, and Precision values in Figure \ref{fig:simu_real}. JRDNN-KM demonstrates a clear and consistent advantage over competing methods. It achieves the highest ARI for cell subgroup identification and also excels in gene network inference. The performance of all methods generally declines from GSD to scMultiSim-T3 to SERGIO-DSI, indicating that the SERGIO-DSI dataset presents the greatest challenge.  In this most difficult scenario, JRDNN-KM maintains its relative effectiveness. While its ARI (0.834) and F1 score (0.481) on SERGIO-DSI are lower than its results on the other datasets, it maintains a similar performance margin over the alternatives. For instance, its lead in F1 score over the second-best method, BLGGM, remains approximately 0.027, 0.027, and 0.024 on GSD, scMultiSim-T3, and SERGIO-DSI, respectively. The observed consistent advantage confirms that the joint learning of subgroups and networks in JRDNN-KM remains effective even in complex and noisy data environments.

\begin{figure*}[!ht]
	\centering
	\includegraphics[scale=0.4]{pic/Complex_Generative.png}
	\caption{Comparative performance of different methods on three application-based simulation datasets; (A)-(D): ARI, F1, Recall, and Precision values.}
	\label{fig:simu_real}
\end{figure*}

We also perform a comprehensive set of simulation studies (Supplementary Section S7) to assess method performance under diverse conditions, including variations in network structure, nonlinearity, common information, dropouts, and sample distributions. JRDNN-KM again proves superior, outperforming other methods in accurately identifying cell subgroups and estimating gene networks.

}

\section{Discussion}
\label{sec:conc}
Estimating heterogeneous gene networks remains a critically important research focus in biomedical science \citep{yang2025ajgm}. In this study, we have proposed JRDNN-KM, a systemic statistical framework for estimating heterogeneous networks and cell subgroups from single-cell transcriptomic data. JRDNN-KM leverages the robust fitting capabilities of neural networks to approximate complex nonlinear gene-gene relationships and integrates the inferred networks into K-means clustering, thereby accounting for heterogeneity in both gene expression levels and gene network structures across cell subgroups. JRDNN-KM has directly addressed the challenging characteristics of single-cell transcriptomic data, including the nonlinearity, zero-inflation, unknown cellular heterogeneity, as well as a certain degree of homogeneity among cell subgroups.

Our analyses of real single-cell transcriptomic data from various sources have validated the utility of JRDNN-KM. Besides the accurate cell subgroup identification performance, the networks identified by JRDNN-KM have been consistently corroborated by biological research. We have analyzed LUAD data as an illustrative case, performing downstream hub gene and community detection analyses based on the estimated networks, which are two pivotal network topological features. Subsequent GO term enrichment analysis has been applied to the two largest communities. As shown in Supplementary Tables S4-S7, our method has uniquely revealed functionally coherent, literature-validated targets such as FTL in ferroptosis regulation and GO:1903320 in regulation of protein modification, which were overlooked by alternative approaches. They offer new insights into homogeneous gene dependencies within tissues alongside subgroup-specific cellular mechanisms governing distinct biological states. As candidate biomarkers, these hub genes and GO term-associated biological functions potentially facilitate target discovery and therapeutic hypothesis generation. Extensive simulation studies have further demonstrated the substantial improvements of JRDNN-KM over recent network estimation models, particularly in handling complex nonlinear relationships among genes that are not adequately captured by models such as GGMs.

In our study, the proposed loss function is based on squared loss, which may lack robustness against outliers and noise. Future work could explore optimizing this aspect to enhance the model's stability and performance under varying conditions. Our analysis utilizes zero-inflated models in accordance with established methods. Alternative approaches employing Poisson or negative binomial distributions, particularly when integrated with deep neural network architectures, represent a compelling avenue for future methodological innovation. Additionally, our assumptions regarding zero inflation are based on varying zero-inflation rates across different genes. However, emerging research indicates that technical dropout may be correlated with expression levels, suggesting a need for a more nuanced approach to account for these dependencies in our model. In heterogeneous analysis, the indicator matrix $\boldsymbol{\Theta}_k$ is used within the Mahalanobis distance framework. Our study is designed to detect nonlinear relationships between variables rather than quantify their precise strength. Consistent with most literature on DNNs, precisely estimating effect magnitudes remains challenging due to the inherent ``black-box'' nature of deep architectures. Future work could further explore quantifying the strength of nonlinear relationships. Our method focuses on point estimates of network structure, consistent with standard practice in the literature, including graphical lasso \citep{zhang2014sparse, dong2023joint}, neighborhood selection \citep{ma2016joint}, and deep learning approaches \citep{shu2021modeling}. While uncertainty quantification would be valuable, it remains challenging for high-dimensional nonlinear frameworks like ours and is deferred to future study. To enhance computational efficiency, we adopt the shared parameter framework, thereby avoiding estimation of nuisance parameters $\sigma_{kj}^2$. This approach has demonstrated satisfactory performance in both simulation studies and real data analyses. Future work could explore estimating $\sigma_{kj}^2$ to enable heterogeneous penalty parameterization.

{In real data analysis, for network validation, we have employed four established reference networks. Although ChIP-seq data provide valuable evidence of direct TF-DNA binding, its utility in our 100-gene context is limited by the sparse representation of TF: too few transcription factors with validated targets were included to form meaningful regulatory connections. Here, we have focused on 100 HV genes, a common practice in existing network estimation studies. Given that we need to consider relationships among $p$ variables (i.e., $p\times (p-1)$ potential edges), an excessively large $p$ would lead to high computational complexity. This has also been a challenging issue in traditional GGMs. In future work, we will focus on optimizing the algorithm to improve model scalability.}

Overall, JRDNN-KM represents a significant advancement in the field of single-cell genomics, offering a powerful tool for researchers exploring the intricate landscape of gene dependence and interactions at the single-cell level.

\section*{Supplementary Materials}

In supplementary materials, the pdf file contains supplementary sections, supplementary figures, and supplementary tables referred in the paper. The zipped file includes the code and datasets used in the simulation and real application and provides detailed instructions for reproducing results. The code that implements the proposed approach is also available at \url{https://github.com/mengyunwu2020/JRDNN-KM}. 

\section*{Acknowledgments}
The authors thank the editors and reviewers for their invaluable feedback and insightful suggestions, which have significantly improved this paper.

\section*{Disclosure Statement}
The authors report there are no competing interests to declare.

\section*{Data Availability Statement}
The data that support the findings of this study are openly available in \url{https://github.com/mengyunwu2020/JRDNN-KM}.

\section*{Funding}
This research was supported by the National Natural Science Foundation of China (12071273); MOE Project of Humanities and Social Sciences (25YJCZH291); Shanghai Rising-Star Program (22QA1403500); Shanghai Science and Technology Development Funds (23JC1402100); Shanghai Research Center for Data Science and Decision Technology; National Institutes of Health (CA204120 and CA121974); and National Science Foundation (2209685).

\bibliographystyle{agsm}

\bibliography{Bibliography-MM-MC}

@article{kiselev2017sc3,
  title={{SC3: consensus clustering of single-cell RNA-seq data}},
  author={Kiselev, Vladimir Yu and Kirschner, Kristina and Schaub, Michael T and Andrews, Tallulah and Yiu, Andrew and Chandra, Tamir and Natarajan, Kedar N and Reik, Wolf and Barahona, Mauricio and Green, Anthony R and others},
  journal={Nature Methods},
  volume={14},
  number={5},
  pages={483--486},
  year={2017},
  publisher={Nature Publishing Group US New York}
}

@article{butler2018integrating,
  title={Integrating single-cell transcriptomic data across different conditions, technologies, and species},
  author={Butler, Andrew and Hoffman, Paul and Smibert, Peter and Papalexi, Efthymia and Satija, Rahul},
  journal={Nature Biotechnology},
  volume={36},
  number={5},
  pages={411--420},
  year={2018},
  publisher={Nature Publishing Group US New York}
}

@article{costanzo2019global,
  title={Global genetic networks and the genotype-to-phenotype relationship},
  author={Costanzo, Michael and Kuzmin, Elena and van Leeuwen, Jolanda and Mair, Barbara and Moffat, Jason and Boone, Charles and Andrews, Brenda},
  journal={Cell},
  volume={177},
  number={1},
  pages={85--100},
  year={2019},
  publisher={Elsevier}
}

@article{zhang2014sparse,
  title={{Sparse precision matrix estimation via lasso penalized D-trace loss}},
  author={Zhang, Teng and Zou, Hui},
  journal={Biometrika},
  volume={101},
  number={1},
  pages={103--120},
  year={2014},
  publisher={Oxford University Press}
}

@article{wang2019precision,
  title={Precision Lasso: accounting for correlations and linear dependencies in high-dimensional genomic data},
  author={Wang, Haohan and Lengerich, Benjamin J and Aragam, Bryon and Xing, Eric P},
  journal={Bioinformatics},
  volume={35},
  number={7},
  pages={1181--1187},
  year={2019},
  publisher={Oxford University Press}
}

@article{cai2016estimating,
  title={Estimating sparse precision matrix: Optimal rates of convergence and adaptive estimation},
  author={Cai, T Tony and Liu, Weidong and Zhou, Harrison H},
  journal={The Annals of Statistics},
  volume={44},
  pages={455--488},
  year={2016},
  publisher={JSTOR}
}

@article{gawad2016single,
  title={Single-cell genome sequencing: current state of the science},
  author={Gawad, Charles and Koh, Winston and Quake, Stephen R},
  journal={Nature Reviews Genetics},
  volume={17},
  number={3},
  pages={175--188},
  year={2016},
  publisher={Nature Publishing Group UK London}
}

@article{mcdavid2019graphical,
  title={Graphical models for zero-inflated single cell gene expression},
  author={McDavid, Andrew and Gottardo, Raphael and Simon, Noah and Drton, Mathias},
  journal={The Annals of Applied Statistics},
  volume={13},
  number={2},
  pages={848--873},
  year={2019},
  publisher={NIH Public Access}
}

@article{xiao2022estimating,
  title={Estimating graphical models for count data with applications to single-cell gene network},
  author={Xiao, Feiyi and Tang, Junjie and Fang, Huaying and Xi, Ruibin},
  journal={Advances in Neural Information Processing Systems},
  volume={35},
  pages={29038--29050},
  year={2022}
}

@article{dong2023joint,
  title={{Joint gene network construction by single-cell RNA sequencing data}},
  author={Dong, Meichen and He, Yiping and Jiang, Yuchao and Zou, Fei},
  journal={Biometrics},
  volume={79},
  number={2},
  pages={915--925},
  year={2023},
  publisher={Wiley Online Library}
}

@article{wang2021constructing,
  title={Constructing local cell-specific networks from single-cell data},
  author={Wang, Xuran and Choi, David and Roeder, Kathryn},
  journal={Proceedings of the National Academy of Sciences},
  volume={118},
  number={51},
  pages={e2113178118},
  year={2021},
  publisher={National Acad Sciences}
}

@article{wu2022estimating,
  title={Estimating heterogeneous gene regulatory networks from zero-inflated single-cell expression data},
  author={Wu, Qiuyu and Luo, Xiangyu},
  journal={The Annals of Applied Statistics},
  volume={16},
  number={4},
  pages={2183--2200},
  year={2022},
  publisher={Institute of Mathematical Statistics}
}

@article{huynh2010inferring,
  title={Inferring regulatory networks from expression data using tree-based methods},
  author={Huynh-Thu, V{\^a}n Anh and Irrthum, Alexandre and Wehenkel, Louis and Geurts, Pierre},
  journal={PloS One},
  volume={5},
  number={9},
  pages={e12776},
  year={2010},
  publisher={Public Library of Science San Francisco, USA}
}

@article{ma2016joint,
  title={Joint structural estimation of multiple graphical models},
  author={Ma, Jing and Michailidis, George},
  journal={The Journal of Machine Learning Research},
  volume={17},
  number={1},
  pages={5777--5824},
  year={2016},
  publisher={JMLR. org}
}

@article{wu2020joint,
  title={Joint learning of multiple gene networks from single-cell gene expression data},
  author={Wu, Nuosi and Yin, Fu and Ou-Yang, Le and Zhu, Zexuan and Xie, Weixin},
  journal={Computational and Structural Biotechnology Journal},
  volume={18},
  pages={2583--2595},
  year={2020},
  publisher={Elsevier}
}

@article{kolodziejczyk2015single,
  title={{Single cell RNA-sequencing of pluripotent states unlocks modular transcriptional variation}},
  author={Kolodziejczyk, Aleksandra A and Kim, Jong Kyoung and Tsang, Jason CH and Ilicic, Tomislav and Henriksson, Johan and Natarajan, Kedar N and Tuck, Alex C and Gao, Xuefei and B{\"u}hler, Marc and Liu, Pentao and others},
  journal={Cell Stem Cell},
  volume={17},
  number={4},
  pages={471--485},
  year={2015},
  publisher={Elsevier}
}

@article{tian2019benchmarking,
  title={{Benchmarking single cell RNA-sequencing analysis pipelines using mixture control experiments}},
  author={Tian, Luyi and Dong, Xueyi and Freytag, Saskia and L{\^e} Cao, Kim-Anh and Su, Shian and JalalAbadi, Abolfazl and Amann-Zalcenstein, Daniela and Weber, Tom S and Seidi, Azadeh and Jabbari, Jafar S and others},
  journal={Nature Methods},
  volume={16},
  number={6},
  pages={479--487},
  year={2019},
  publisher={Nature Publishing Group US New York}
}

@article{zheng2017massively,
  title={Massively parallel digital transcriptional profiling of single cells},
  author={Zheng, Grace XY and Terry, Jessica M and Belgrader, Phillip and Ryvkin, Paul and Bent, Zachary W and Wilson, Ryan and Ziraldo, Solongo B and Wheeler, Tobias D and McDermott, Geoff P and Zhu, Junjie and others},
  journal={Nature Communications},
  volume={8},
  number={1},
  pages={14049},
  year={2017},
  publisher={Nature Publishing Group UK London}
}

@article{han2018mapping,
  title={Mapping the mouse cell atlas by microwell-seq},
  author={Han, Xiaoping and Wang, Renying and Zhou, Yincong and Fei, Lijiang and Sun, Huiyu and Lai, Shujing and Saadatpour, Assieh and Zhou, Ziming and Chen, Haide and Ye, Fang and others},
  journal={Cell},
  volume={172},
  number={5},
  pages={1091--1107},
  year={2018},
  publisher={Elsevier}
}

@article{haghverdi2018batch,
  title={{Batch effects in single-cell RNA-sequencing data are corrected by matching mutual nearest neighbors}},
  author={Haghverdi, Laleh and Lun, Aaron TL and Morgan, Michael D and Marioni, John C},
  journal={Nature Biotechnology},
  volume={36},
  number={5},
  pages={421--427},
  year={2018},
  publisher={Nature Publishing Group}
}

@article{cole2019performance,
  title={{Performance assessment and selection of normalization procedures for single-cell RNA-seq}},
  author={Cole, Michael B and Risso, Davide and Wagner, Allon and DeTomaso, David and Ngai, John and Purdom, Elizabeth and Dudoit, Sandrine and Yosef, Nir},
  journal={Cell Systems},
  volume={8},
  number={4},
  pages={315--328},
  year={2019},
  publisher={Elsevier}
}

@article{booeshaghi2021normalization,
  title={{Normalization of single-cell RNA-seq counts by log(x+ 1) or log(1+ x)}},
  author={Booeshaghi, A Sina and Pachter, Lior},
  journal={Bioinformatics},
  volume={37},
  number={15},
  pages={2223--2224},
  year={2021},
  publisher={Oxford University Press}
}

@article{hafemeister2019normalization,
  title={{Normalization and variance stabilization of single-cell RNA-seq data using regularized negative binomial regression}},
  author={Hafemeister, Christoph and Satija, Rahul},
  journal={Genome Biology},
  volume={20},
  number={1},
  pages={296},
  year={2019},
  publisher={Springer}
}

@article{li2023zinbmm,
  title={{ZINBMM: a general mixture model for simultaneous clustering and gene selection using single-cell transcriptomic data}},
  author={Li, Yang and Wu, Mingcong and Ma, Shuangge and Wu, Mengyun},
  journal={Genome Biology},
  volume={24},
  number={1},
  pages={208},
  year={2023},
  publisher={Springer}
}

@article{oh2023accounting,
  title={{Accounting for technical noise in Bayesian graphical models of single-cell RNA-sequencing data}},
  author={Oh, Jihwan and Chang, Changgee and Long, Qi},
  journal={Biostatistics},
  volume={24},
  number={1},
  pages={161--176},
  year={2023},
  publisher={Oxford University Press}
}

@article{nguyen2023structure,
  title={{Structure learning for zero-inflated counts with an application to single-cell RNA sequencing data}},
  author={Nguyen, Thi Kim Hue and Van den Berge, Koen and Chiogna, Monica and Risso, Davide},
  journal={The Annals of Applied Statistics},
  volume={17},
  number={3},
  pages={2555--2573},
  year={2023},
  publisher={Institute of Mathematical Statistics}
}

@article{karaaslanli2023kernelized,
  title={{Kernelized multiview signed graph learning for single-cell RNA sequencing data}},
  author={Karaaslanli, Abdullah and Saha, Satabdi and Maiti, Tapabrata and Aviyente, Selin},
  journal={BMC Bioinformatics},
  volume={24},
  number={1},
  pages={127},
  year={2023},
  publisher={Springer}
}

@article{halama2024roadmap,
  title={A roadmap to the molecular human linking multiomics with population traits and diabetes subtypes},
  author={Halama, Anna and Zaghlool, Shaza and Thareja, Gaurav and Kader, Sara and Al Muftah, Wadha and Mook-Kanamori, Marjonneke and Sarwath, Hina and Mohamoud, Yasmin Ali and Stephan, Nisha and Ameling, Sabine and others},
  journal={Nature Communications},
  volume={15},
  number={1},
  pages={7111},
  year={2024},
  publisher={Nature Publishing Group UK London}
}

@article{wang2022addressing,
  title={Addressing the mean-correlation relationship in co-expression analysis},
  author={Wang, Yi and Hicks, Stephanie C and Hansen, Kasper D},
  journal={PLoS Computational Biology},
  volume={18},
  number={3},
  pages={e1009954},
  year={2022},
  publisher={Public Library of Science San Francisco, CA USA}
}

@article{wang2021single,
  title={Single-cell normalization and association testing unifying CRISPR screen and gene co-expression analyses with Normalisr},
  author={Wang, Lingfei},
  journal={Nature Communications},
  volume={12},
  number={1},
  pages={6395},
  year={2021},
  publisher={Nature Publishing Group UK London}
}

@article{su2023cell,
  title={Cell-type-specific co-expression inference from single cell RNA-sequencing data},
  author={Su, Chang and Xu, Zichun and Shan, Xinning and Cai, Biao and Zhao, Hongyu and Zhang, Jingfei},
  journal={Nature Communications},
  volume={14},
  number={1},
  pages={4846},
  year={2023},
  publisher={Nature Publishing Group UK London}
}

@article{yang2025ajgm,
  title={{AJGM: joint learning of heterogeneous gene networks with adaptive graphical model}},
  author={Yang, Shunqi and Hu, Lingyi and Chen, Pengzhou and Zeng, Xiangxiang and Mao, Shanjun},
  journal={Bioinformatics},
  volume={41},
  number={3},
  pages={btaf096},
  year={2025},
  publisher={Oxford University Press}
}

@article{shu2021modeling,
  title={Modeling gene regulatory networks using neural network architectures},
  author={Shu, Hantao and Zhou, Jingtian and Lian, Qiuyu and Li, Han and Zhao, Dan and Zeng, Jianyang and Ma, Jianzhu},
  journal={Nature Computational Science},
  volume={1},
  number={7},
  pages={491--501},
  year={2021},
  publisher={Nature Publishing Group US New York}
}

@article{pratapa2020benchmarking,
  title={Benchmarking algorithms for gene regulatory network inference from single-cell transcriptomic data},
  author={Pratapa, Aditya and Jalihal, Amogh P and Law, Jonathan N and Bharadwaj, Aditya and Murali, TM},
  journal={Nature Methods},
  volume={17},
  number={2},
  pages={147--154},
  year={2020},
  publisher={Nature Publishing Group},
  doi={10.1038/s41592-019-0690-6}
}

@article{li2025scmultisim,
  title={{scMultiSim: simulation of single-cell multi-omics and spatial data guided by gene regulatory networks and cell--cell interactions}},
  author={Li, Hechen and Zhang, Ziqi and Squires, Michael and Chen, Xi and Zhang, Xiuwei},
  journal={Nature Methods},
  pages={982--993},
  volume={22},
  year={2025},
  publisher={Nature Publishing Group US New York}
}

@article{dibaeinia2020sergio,
  title={{SERGIO: a single-cell expression simulator guided by gene regulatory networks}},
  author={Dibaeinia, Payam and Sinha, Saurabh},
  journal={Cell Systems},
  volume={11},
  number={3},
  pages={252--271},
  year={2020},
  publisher={Elsevier}
}

@article{chevalley2025large,
  title={A large-scale benchmark for network inference from single-cell perturbation data},
  author={Chevalley, Mathieu and Roohani, Yusuf H and Mehrjou, Arash and Leskovec, Jure and Schwab, Patrick},
  journal={Communications Biology},
  volume={8},
  number={1},
  pages={412},
  year={2025},
  publisher={Nature Publishing Group UK London}
}
\end{document}